# Graph Neural Network Prediction of Frequency Nadir and RoCoF Under Topology Variations

Linhan Fang
Department of Electrical and Computer Engineering
University of Houston
Houston, TX, USA
lfang7@uh.edu

Fan Jiang
Department of Electrical and Computer Engineering
University of Houston
Houston, TX, USA
fjiang6@uh.edu

Xingpeng Li
Department of Electrical and Computer Engineering
University of Houston
Houston, TX, USA
xli82@uh.edu

***Abstract*—Accurate prediction of frequency dynamics under varying network topologies is essential for rapid frequency security assessment, whereas repeated electromagnetic transient (EMT) simulations are computationally intensive. This paper develops a topology-aware graph neural network with a radial basis function Kolmogorov–Arnold network (RBF-KAN) regression head, termed RK-GNN, to predict frequency nadir and maximum absolute rate of change of frequency (RoCoF). Scenarios combine load scaling, regional load redistribution, and generator redispatch with intact, N−1, and N−2 transmission topologies. After checks for network connectivity, generator reference limits, and numerical consistency, the retained scenarios are simulated in PSCAD to construct the reference dataset. The RK-GNN is evaluated against a conventional deep neural network (DNN) and a baseline GNN. On the IEEE 39-bus system, the RK-GNN reduces the mean absolute errors (MAEs) of nadir and RoCoF by 48% and 59% relative to the DNN and by 42% and 61% relative to the baseline GNN. Incorporating the bridge indicator and disturbance hop distance yields further MAE reductions of 27% and 54%. These results show that the proposed topology-aware architecture provides accurate prediction of both frequency metrics under varying topologies and substantially reduces the need for repeated EMT simulations.**

***Index Terms*—Electromagnetic transient simulation, Frequency nadir, Frequency security assessment, Graph neural network, Rate of change of frequency, Topology variation.**

## I. INTRODUCTION

A sudden generator outage creates an active power imbalance that causes system frequency to decline. The severity of the resulting frequency response depends on the size of the outage, system inertia, primary frequency response, and the pre-disturbance operating condition [1]-[2]. The rate of change of frequency (RoCoF) quantifies the rate of frequency deviation immediately after a disturbance, whereas the frequency nadir denotes the minimum frequency reached during the subsequent response [1]. The maximum RoCoF and frequency nadir are commonly used to characterize short-term frequency security and assess system response to generator outages.

Time-domain simulation is widely used to examine system dynamics following disturbances [3]. Electromagnetic transient (EMT) simulation captures detailed interactions among generators, control systems, and network components, and platforms such as PSCAD/EMTDC are commonly used for detailed dynamic studies [3]-[4]. This modeling fidelity, however, entails substantial computational cost because small integration time steps are required and each operating condition, network configuration, and contingency must be simulated separately over the full simulation horizon [4]-[5]. As the number of scenarios increases, repeated EMT simulations become increasingly expensive, limiting their practicality for large-scale frequency security assessment and motivating the use of faster prediction models.

To reduce the computational burden of repeated dynamic simulations, machine learning (ML) models have been investigated for frequency security assessment. Artificial neural networks have been used to predict frequency nadir and RoCoF from pre-disturbance operating information following generator outages [6], while other studies have compared different learning algorithms for rapid frequency nadir estimation [7]. Deep neural networks (DNNs) have also been applied to estimate locational RoCoF under varying operating conditions [8] and, more recently, to jointly predict RoCoF and frequency nadir for operations subject to frequency constraints [9]. These studies demonstrate that data-driven models can estimate key frequency metrics with substantially lower online computational cost. However, most existing predictors are developed using operating conditions drawn from a limited set of network configurations, leaving their performance under topology changes less well characterized.

In practical operation, network topology can change due to scheduled maintenance, line outages, or prior contingencies. These changes alter power transfer paths and redistribute pre-disturbance power flows [10], such that similar load and generation conditions can lead to different dynamic responses. Previous studies have examined ML-based stability assessment across multiple transmission topologies [11]-[12], while more recent work has investigated representations designed to improve generalization to previously unseen configurations [13]. Nevertheless, conventional vector-based DNNs typically encode bus, generator, and line-status variables as flattened, fixed-length inputs. Although topology information can be included, such representations do not explicitly preserve bus-line connectivity or local electrical relationships. Their prediction accuracy can become sensitive to topologies that are poorly represented or entirely absent from the training data.

Graph neural networks (GNNs) provide a direct way to address this limitation by representing the power system as a graph and incorporating network connectivity into the learning process [14]. Recent studies have applied graph-based models to frequency health prediction [15] and transient stability assessment [16], demonstrating their ability to capture spatial dependencies and electrical relationships across the network.

Accordingly, this paper develops a topology-aware GNN model to predict frequency nadir and maximum RoCoF following generator outages using information available before EMT simulation. Training data are generated from PSCAD simulations of screened scenarios covering varying operating conditions and intact, *N*−1, and *N*−2 transmission topologies. The proposed GNN is compared with a conventional DNN and a baseline GNN in terms of prediction accuracy.

## II. METHODOLOGY

This section presents a data-driven framework for predicting post-disturbance frequency dynamics under varying network topologies as shown in Fig. 1. Operating conditions and topology scenarios are first constructed and screened, followed by EMT simulations to obtain the corresponding frequency responses and extract the target frequency metrics. The resulting dataset is then used to train a topology-aware graph neural network enhanced with a Kolmogorov–Arnold network (KAN) [17] regression head, in which the B-spline basis is replaced by Gaussian radial basis functions [18] for computational efficiency; the resulting model is referred to as RK-GNN. A conventional DNN and a baseline GNN are adopted for performance comparison.

### A. Scenario Generation and Screening

The dataset is constructed on the IEEE 39-bus system by varying both operating conditions and network topology. Operating variations include changes in the overall loading level, regional load distribution, and generator dispatch. Uniform load scaling modifies the system loading level while proportionally adjusting generator active power references. Regional load variations redistribute demand among different areas while maintaining the total system load, whereas generator redispatch changes the spatial distribution of generation without altering the total generation. Topology variations are introduced through the intact network and connected N−1 and N−2 configurations obtained by removing one or two transmission lines, respectively. Generator trip events are then sampled across the resulting operating and topology conditions.

The screening stage adopts the standard assumptions of flat voltage magnitudes, small angle differences, and negligible branch resistance. The nodal power balance is defined as:

$$P = B'\theta \tag{1}$$

where $P_i = P_{g,i} - P_{L,i}$ denotes the net active power injection at bus $i$, $\theta$ is the vector of bus voltage angles, $B'$ is the susceptance matrix of the current topology, and $a_{ij}$ is the transformer tap ratio, with $B'_{ij} = -1/(a_{ij}x_{ij}), i \neq j$, and $B'_{ii} = \sum_{j\neq i} 1/(a_{ij}x_{ij})$. The active power flow on branch $ij$ follows from the angle solution as:

$$P_{ij} = (\theta_i - \theta_j)/(a_{ij}x_{ij}) \tag{2}$$

A candidate scenario is retained only if three conditions hold. First, the graph formed by the remaining branches must be connected, which guarantees that the reduced matrix $B'$ obtained after removing the slack bus is invertible and that (1) admits a unique solution. Second, the generator active power references must satisfy (3). The residual mismatch is assigned to the reference bus only to balance (1), and the generator references passed to the EMT model are not modified.

$$P_{g,i}^{min} \leq P_{g,i} \leq P_{g,i}^{max} \tag{3}$$

Third, the residual of the solution must satisfy $\|P - B'\theta\| \leq \varepsilon$ to exclude numerically unreliable cases. A scenario is discarded if any of these conditions is violated. For the retained scenarios, the resulting bus voltage angles, branch active power flows, and line loadings are stored as diagnostic quantities to characterize the pre-disturbance network condition.

### B. EMT Simulation and Target Extraction

The screened scenarios are evaluated using EMT simulation in PSCAD to obtain the reference frequency response for each case. The EMT model represents the detailed dynamics of synchronous machines, excitation systems, governor controls, and network components using a small simulation time step. This enables the transient frequency response following a generator trip to be captured under the operating conditions and network topologies defined in the previous stage. Each simulation is specified by four groups of scenario inputs. The bus loads $P_L$ and $Q_L$, together with the generator dispatch $P_g$, define the pre-disturbance operating point. The line status vector defines the network topology, while the generator trip event specifies the unit disconnected at time $t_0$ after the system reaches steady state.

Two scalar frequency metrics are extracted within an observation window $T$ = 2.5 s following the disturbance. The frequency of each online generator is obtained from the PSCAD rotor-speed output at $\Delta t$ = 0.25 ms. RoCoF is computed as an unfiltered two-point finite difference over a 166.75-ms sliding window advanced in 10-ms steps. Let $f_i(t)$ denote the frequency of online generator $i$. The frequency nadir and maximum absolute RoCoF are expressed as:

$$f_{nadir} = \min_{i\in\Omega} \min_{t\in[t_0,t_0+T]} f_i(t) \tag{4}$$

$$RoCoF_{max} = \max_{i\in\Omega} \max_{t\in[t_0,t_0+T]} \left|\frac{df_i(t)}{dt}\right| \tag{5}$$

where Ω denotes the set of generators that remain online after the trip. These two quantities characterize the post-disturbance frequency response and are used as the prediction targets for the ML models.

Since each EMT run is computationally intensive, the simulations are used only to construct a representative reference dataset. A data-driven model is then trained on this dataset to predict the frequency nadir and maximum absolute RoCoF directly, without repeating the full EMT simulation for each new scenario.

### C. ML Models for Frequency Response Prediction

The prediction models are designed to estimate the frequency metrics obtained from PSCAD without repeating the full EMT simulation for each new scenario. Therefore, the inputs are restricted to information available before the disturbance and used to configure the corresponding EMT case. They consist of four blocks: bus loads $\mathbf{x}_i = [P_{L,i}, Q_{L,i}]$; generator active power reference, inertia constant, machine rating, and trip indicator; branch parameters $r_{ij}$, $x_{ij}$, $b_{ij}$ and transformer tap ratio; and the branch status $s_{ij}$, which indicates whether branch $ij$ is in service. The prediction task is expressed as:

$$\hat{y} = \begin{bmatrix} \hat{f}_{nadir} \\ \widehat{RoCoF}_{max} \end{bmatrix} = F_\Theta(\mathbf{x}, \mathbf{g}, \mathbf{e}, \mathbf{s}) \tag{6}$$

where $\mathbf{x}, \mathbf{g}, \mathbf{e}, \mathbf{s}$ denote the four input blocks and Θ represents the trainable parameters.

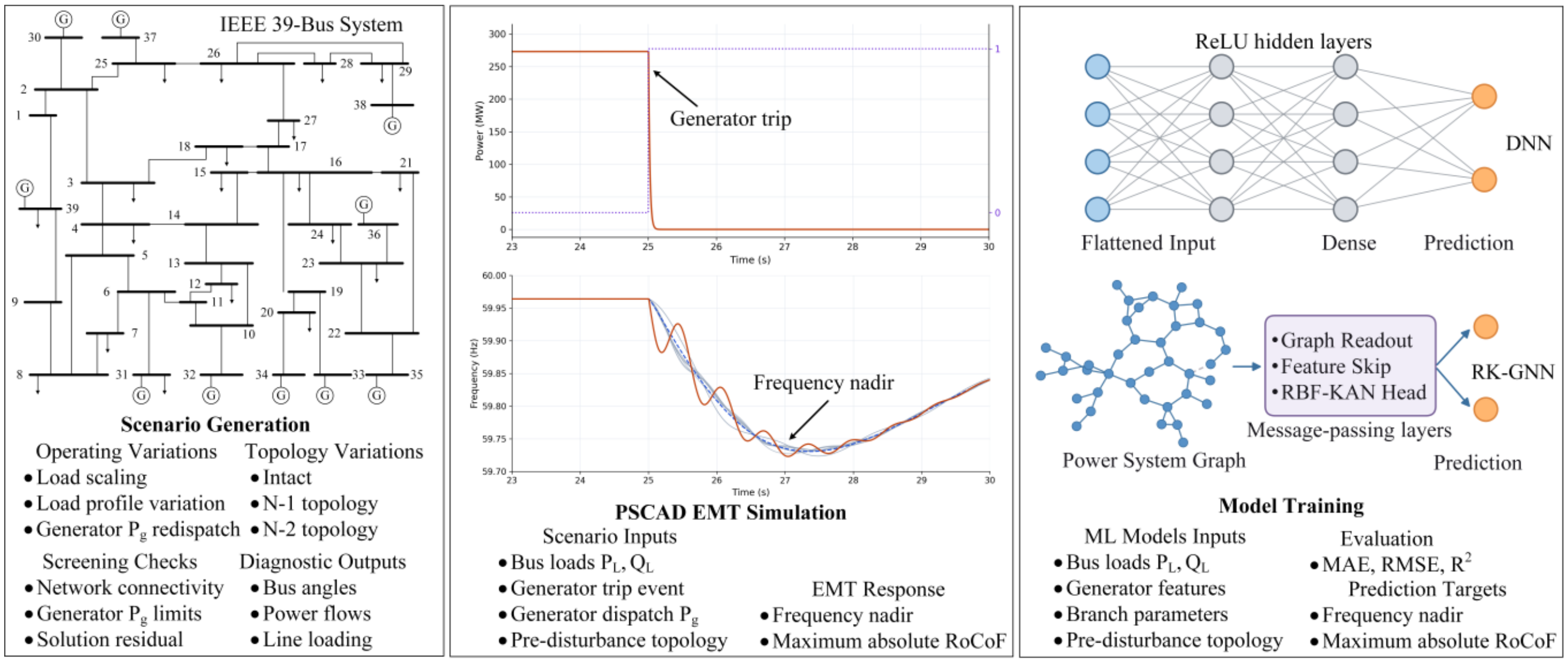

Fig. 1: Overall framework of the proposed frequency-security prediction methodology.

Two reference models are considered for comparison. The conventional DNN flattens the four input blocks together with the fixed network mappings into a single vector, which is processed by fully connected layers with ReLU activation and a linear output layer. Although branch status is included, the flattened representation does not explicitly preserve graph locality or exchange information according to network connectivity. The baseline GNN represents buses as nodes and branches as edges, with bus, generator, and branch features encoded separately. Generator embeddings are mapped to their connected buses, and edge-aware message passing is performed over the in-service branches. The resulting bus and generator representations are aggregated through mean pooling and processed by a conventional dense regression head.

The proposed RK-GNN uses the same graph representation, with buses as nodes and branches as edges. Bus, generator, and branch features are encoded separately, and generator embeddings are mapped to their connected buses. Branch status determines the active graph and therefore directly controls the paths of message propagation. For bus $i$, information from its in-service neighbors is aggregated as:

$$\bar{\mathbf{m}}_i^{(l)} = \frac{1}{|\mathcal{N}_i|} \sum_{j \in \mathcal{N}_i} \phi\left(\mathbf{h}_i^{(l)}, \mathbf{h}_j^{(l)}, \mathbf{h}_{ij}^{e}\right) \tag{7}$$

The node representation is updated by:

$$\mathbf{h}_i^{(l+1)} = \mathbf{h}_i^{(l)} + \alpha\psi\left(\mathbf{h}_i^{(l)}, \bar{\mathbf{m}}_i^{(l)}\right) \tag{8}$$

where $\mathcal{N}_i$ is the set of in-service neighbors, $\mathbf{h}_i^{(l)}$ is the bus representation at layer $l$, $\mathbf{h}_{ij}^{e}$ is the encoded branch feature, and $\phi$ and $\psi$ are dense transformations with SiLU activation, and $\alpha$ is the residual scaling factor. After two message passing layers, the readout combines the final bus states, generator embeddings, the tripped generator and its connected bus representation, and a skip connection containing the original input blocks.

The combined representation is reduced through a linear bottleneck and passed to an RBF-KAN regression head. Its non-linear mapping incorporates both a SiLU-based projection and Gaussian radial basis functions, as formulated below:

$$\varphi(z) = w^{(b)}\mathrm{SiLU}(z) + \sum_{c=1}^{C} w_c \exp\left[-\left(\frac{z-\mu_c}{\sigma}\right)^2\right] \tag{9}$$

where $z$ is the scalar layer input, $C$ is the number of Gaussian basis functions, $\mu_c$ and $\sigma$ denote the basis center and width, and $w^{(b)}$ and $w_c$ are trainable coefficients.

The final layer outputs the predicted frequency nadir and maximum absolute RoCoF. Inputs and targets are standardized using the training set, and the model is optimized using a weighted mean squared error to balance the two targets.

## III. Case Studies

This section evaluates the proposed framework on the IEEE 39-bus system [19]. The dataset construction, data partition, and training settings of all models are described. The RK-GNN is compared with the DNN and baseline GNN across intact, N−1, and N−2 topologies in terms of prediction accuracy. Finally, the effect of appending features derived from network science to the RK-GNN inputs is examined.

### *A. Model Implementation and Training*

The dataset consists of 545 randomly generated generator-trip cases, comprising 66 intact, 462 N−1, and 17 N−2 cases. The samples are split into training, validation, and test sets, with all cases sharing the same pre-disturbance topology and operating point kept in the same subset so that no operating point seen in training appears in validation or testing. The resulting split contains 361 training, 92 validation, and 92 test samples. All models use the same partitions, and inputs and targets are standardized using training-set statistics. The generator at bus 38 is modeled as two identical units on the same bus, giving 11 generating units in total.

The DNN takes a 455-dimensional input formed by 352 system features and 103 fixed network mapping values. Two hidden layers with 64 ReLU units each are followed by a linear output layer, giving 33,474 trainable parameters. The baseline GNN encodes bus, generator, and branch features with dense ReLU layers of 32, 16, and 16 units, respectively. Generator embeddings are merged into their connected buses through a 32-unit layer, after which six residual message passing blocks operate on 32-dimensional node states using branch embeddings, mean aggregation over in-service neighbors, and layer normalization. The final bus and generator states are mean pooled and concatenated, then passed through a 32-unit ReLU layer to a linear output. The model has 32,066 parameters. The RK-GNN is summarized in Table I. Bus, generator, and branch inputs are encoded into 16-dimensional representations, and generator embeddings are fused into their connected buses to form the initial node states for two residual message passing blocks. The readout concatenates the final bus states, generator embeddings, the tripped generator and its bus, and a skip of the raw inputs into a 1,184-dimensional vector, which is compressed to 16 dimensions by a linear bottleneck before the RBF-KAN head. The model has 24,962 parameters, 25.4% fewer than the DNN.

TABLE I. ARCHITECTURE OF THE RK-GNN

| Module | Output shape | Parameters |
|---|---|---|
| Generator encoder | 11×16 | 80 |
| Bus encoder | 39×16 | 48 |
| Branch encoder | 46×16 | 80 |
| Bus-generator fusion | 39×16 | 528 |
| Residual message block 1 | 39×16 | 1,328 |
| Residual message block 2 | 39×16 | 1,328 |
| Readout & input skip | 1,184 | 0 |
| Bottleneck projection | 16 | 18,960 |
| RBF-KAN hidden | 16 | 2,320 |
| RBF-KAN output | 2 | 290 |

The DNN, baseline GNN, and RK-GNN are trained with full-batch Adam optimizer with an initial learning rate of $10^{-3}$, gradient norm clipping applied independently to each parameter tensor at 5, and L2 regularization on the dense-layer kernels. Validation performance is evaluated every five epochs. For the DNN and baseline GNN, the L2 coefficient is $10^{-4}$, and checkpoints are selected according to the validation MSE of the standardized targets, with a minimum improvement threshold of $10^{-5}$. The learning rate is reduced by half after 80 epochs without sufficient validation improvement, down to a minimum of $10^{-5}$. Early stopping uses a patience of 200 epochs, with a maximum training budget of 1,500 epochs. For the proposed RK-GNN model, eight predefined candidates are formed by combining four architectural variants with RoCoF loss weights of 1 and 2. Each candidate is trained for 1,500 epochs using an L2 coefficient of $10^{-5}$ and cosine learning-rate decay from $10^{-3}$ to $5\times10^{-5}$. At each validation checkpoint, the RMSE of each target is normalized by the corresponding fixed DNN validation RMSE. The candidate and checkpoint minimizing the larger normalized error are retained, with the mean normalized error used to resolve ties. The selected RK-GNN assigns a weight of 2 to the RoCoF term in the standardized-target training loss. The test set is excluded from all candidate, hyperparameter, and checkpoint selection.

### *B. Prediction Performance Across Topology Variations*

Fig. 2 compares the DNN, baseline GNN, and RK-GNN across different topology categories. Over all 92 test cases, the RK-GNN achieves nadir and maximum absolute RoCoF MAEs of 0.0059 Hz and 0.0161 Hz/s, corresponding to reductions of approximately 48% and 59% relative to the DNN and 42% and 61% relative to the baseline GNN. Similar improvements are observed in RMSE, while the $R^2$ values reach 0.9982 and 0.9966 for nadir and RoCoF, respectively.

The baseline GNN improves nadir prediction over the DNN but provides no consistent improvement in RoCoF, indicating that graph representation alone does not guarantee higher accuracy for both metrics. Prediction errors generally increase from intact to N−1 and N−2 topologies, reflecting the increased modeling difficulty introduced by topology variations. The RK-GNN provides the lowest overall errors for both frequency metrics and the lowest RoCoF error in every topology category, while the three models yield comparable N−2 nadir errors.

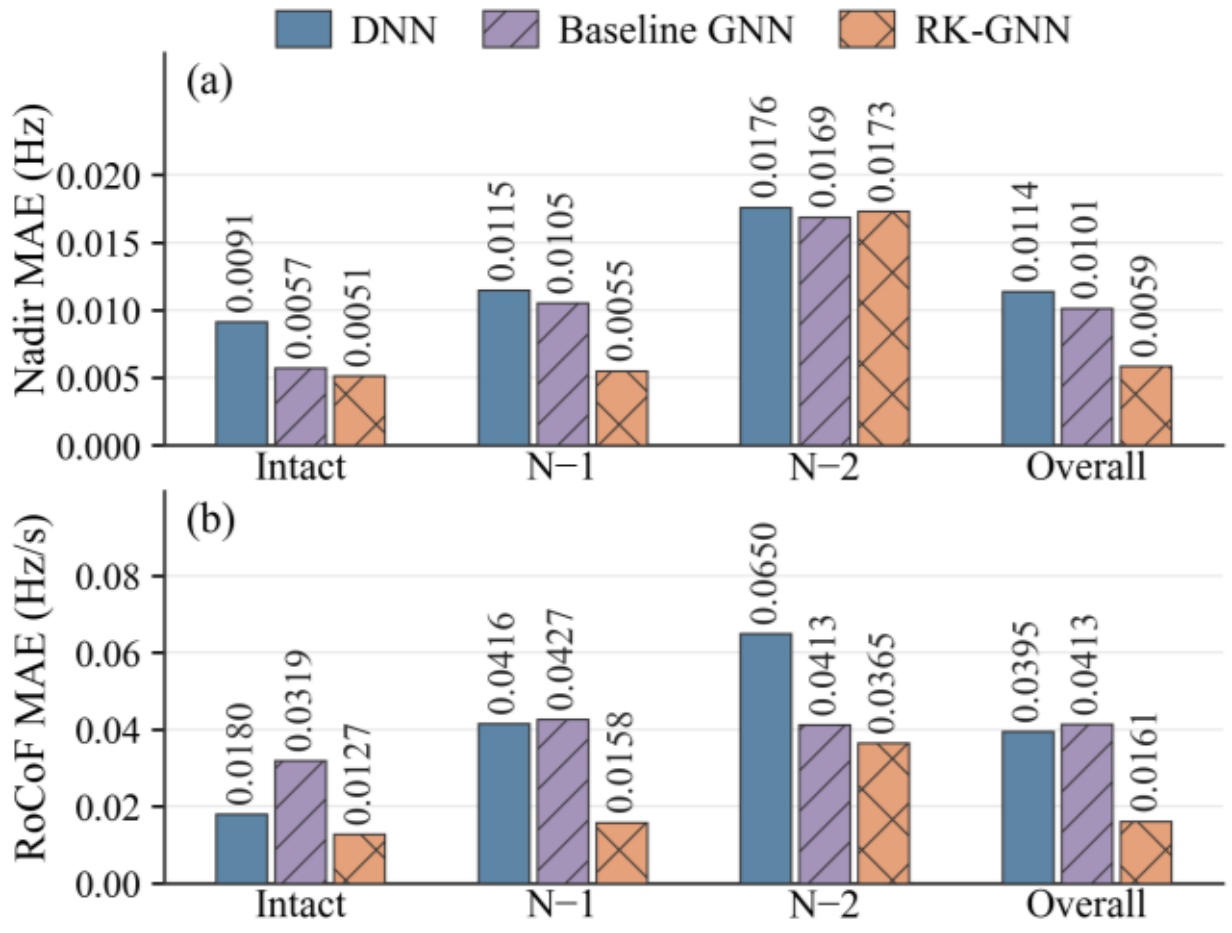

Fig. 2: MAEs of the DNN, baseline GNN, and RK-GNN across different topology categories: (a) frequency nadir; (b) maximum absolute RoCoF.

Fig. 3 further compares the predicted and PSCAD simulated values. The RK-GNN predictions are more tightly concentrated around the ideal line for both targets. In particular, 84% of its nadir predictions have an absolute error below 0.01 Hz, compared with 53% for the DNN, while 95% of its RoCoF predictions remain within 0.05 Hz/s of the PSCAD values, compared with 70% for the DNN. Overall, the RK-GNN achieves the lowest errors on the full test set, with the largest improvement observed for RoCoF.

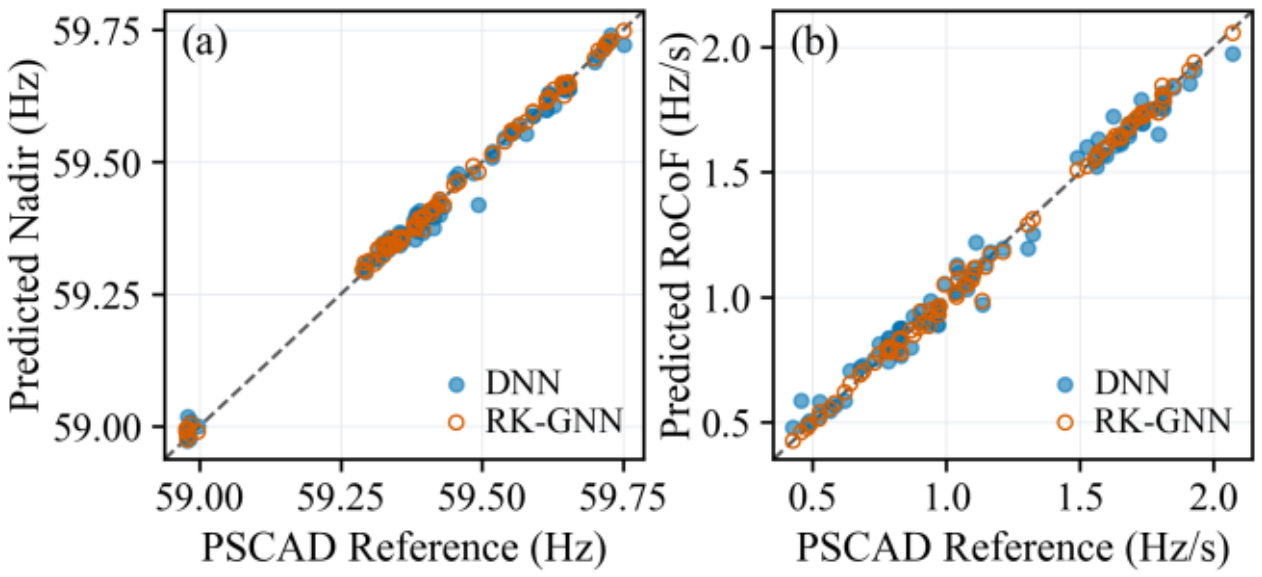

Fig. 3: Test comparison of DNN and RK-GNN predictions against PSCAD-simulated (a) frequency nadir and (b) maximum absolute RoCoF.

Generating the 545-scenario dataset required 27.31 h of PSCAD simulation, or approximately 180 s per case. In contrast, the trained RK-GNN evaluates all scenarios in 0.0987 s for the 545-case batch, corresponding to an average inference time of 0.18 ms per scenario, enabling rapid frequency assessment after offline dataset construction.

### *C. Effect of Network Science Features*

The RK-GNN without graph-derived features relies solely on the physical attributes and connectivity of each operating scenario. With two message passing layers, the representation of each bus incorporates information from nodes within at most two graph hops, whereas the IEEE 39-bus system contains considerably longer topological paths. Although the subsequent readout aggregates system-wide representations, information associated with remote disturbance locations and global connectivity characteristics is not explicitly propagated through the local message passing process. This motivates the introduction of graph-derived features from network science to assess whether such global structural information can complement local message passing for frequency metric prediction.

Seven features drawn from network science and its electrical adaptations are evaluated, including node degree, electrical strength, node betweenness centrality, hop distance (HD), effective distance, bridge indicator (BF), and edge betweenness centrality. BF identifies active branches whose removal disconnects the network, while HD measures the shortest path from each bus to the tripped generator bus. All 128 feature combinations are evaluated while keeping the RK-GNN architecture, training procedure, and data split unchanged. The final subset is selected on the validation set only, and the test set is used solely for the reported evaluation. As shown in Table II, the RK-GNN improves upon the baseline GNN, reducing the nadir and RoCoF MAEs from 0.0101 to 0.0059 Hz and from 0.0413 to 0.0161 Hz/s, respectively. Incorporating the selected BF and HD further reduces these errors to 0.0043 Hz and 0.0074 Hz/s, with corresponding RMSE reductions of approximately 28% and 60% relative to the RK-GNN. These results demonstrate that the proposed architecture provides substantial gains over the baseline GNN, while compact graph-derived features offer additional improvement, particularly for RoCoF prediction.

TABLE II. EFFECT OF SELECTED GRAPH-DERIVED FEATURES ON PREDICTION PERFORMANCE

| Target | Metric | Baseline GNN | RK-GNN | RK-GNN +BF+HD | Δ |
|---|---|---|---|---|---|
| Nadir | MAE (Hz) | 0.0101 | 0.0059 | 0.0043 | -27.2% |
| | RMSE (Hz) | 0.0147 | 0.0080 | 0.0057 | -28.2% |
| RoCoF | MAE (Hz/s) | 0.0413 | 0.0161 | 0.0074 | -54.3% |
| | RMSE (Hz/s) | 0.0583 | 0.0257 | 0.0102 | -60.3% |

## IV. CONCLUSION

This paper presented a topology-aware graph neural network with an RBF-KAN regression head for predicting frequency nadir and maximum absolute RoCoF under varying network topologies. Scenarios spanning load, dispatch, and intact, N−1, and N−2 topology variations were screened and simulated in PSCAD to construct the reference dataset. On the IEEE 39-bus system, the RK-GNN reduced nadir and RoCoF MAEs by approximately 48% and 59% relative to the DNN and by 42% and 61% relative to the baseline GNN, achieving the lowest RoCoF error in every topology category with fewer parameters than either reference model. Incorporating the bridge indicator and hop distance further reduced the MAEs to 0.0043 Hz and 0.0074 Hz/s, showing that compact network science features complement local message passing. Once trained, the model evaluates a scenario in 0.18 ms, compared with about 180 s per EMT simulation. Future work will extend the framework to larger systems and online frequency security assessment.